# WDM NOMA VLC Systems

Mansourah K. A. Aljohani [1], Mohamed O. I. Musa[1], Mohammed T. Alresheedi[2] and Jaafar M. H. Elmirghani[1]

[1]School of Electronic and Electrical Engineering, University of Leeds, Leeds, LS2 9JT, UK
[2]Department of Electrical Engineering, King Saud University, Riyadh, Kingdom of Saudi Arabia
E-mail: {ml16mka, m.musa, j.m.h.elmirghani}@leeds.ac.uk [1]

**ABSTRACT**
Wavelength division multiplexing (WDM) and non-orthogonal multiple access (NOMA) are employed in a visible light communication (VLC) system to increase the capacity of the system. The system is evaluated using two different scenarios focusing on power and data rate, where the first scenario is based on fair power allocation while the second scenario provides equal power allocation. Data rate variation is evaluated as a function of the users positioning and mobility. In both scenarios the proposed NOMA-WDM system achieved higher data rate NOMA system.
**Keywords**: Visible Light communication, Wavelength division multiplexing, Non-Orthogonal VLC, WDM, NOMA, OW.

## 1. INTRODUCTION

Visible light Communication (VLC) and Optical Wireless communication (OWC) have several advantages compared to Radio Frequency (RF) communication. These advantages are mainly focused on the increased security in the infrastructure [1]. The need for (VLC) systems was due to the limited radio spectrum availability, limited channel capacity and the increased demand for data rates by users [2], [3]. Signals in VLC systems are transmitted via transmitters that use Laser Diodes (LD) designed specifically for use in VLC systems to perform electrical-optical conversion. Red, Green, Yellow and Blue (RGYB) LDs, (ie white LDs engines) are attractive due to their ability to their conversion efficiency and high power levels. Nevertheless, they are cost efficient and can work with various power levels. [4]

Studies have been conducted on VLC systems in order to enhance the system performance and increase the achievable data rates and power efficiency [5]-[25]. To provide multiple access, Non-Orthogonal Multiple Access (NOMA) can be used in VLC systems. NOMA operates in a similar fashion to a multiplexer and demultiplexer. It multiplexes all its users' signals in the power domain then demultiplex them at the receiver end. NOMA uses two different methods in data transmission and receiving, for transmission it uses superposition coding and successive interference cancellation (SIC) in the receiver to separate the data streams. NOMA has several benefits over Orthogonal Multiple access techniques. These can be summarised in: (a) In NOMA, the channel gain can be different for different channels. This allows the channels to use the same resources, with SIC separating the data, thus making better use of resources and increasing the overall capacity, potentially. (b) NOMA makes better use of the power domain than other techniques. It allocates power to users according to the level of their channel gain. If the user has high channel gain it is allocated low power and vice versa, which provides better signal processing capabilities. (c) In order for a NOMA user to multiplex it does not need prior information of the instantaneous frequency-selective fading channel conditions, such as the frequency-selective channel quality indicator (CQI) or channel state information (CSI) which are used in the receiver end. Therefore, this can improve the performance gain of NOMA regardless of the users' mobility or the delay caused by CSI feedback,. [5].

In NOMA, the signal power is allocated to users according to their channel strengths. The lower the channel condition the higher the signal power allocated to help users decode their messages. On the other hand, users who experience better channel conditions employ SIC to allocate and decode their messages as they are assigned lower signal power, [6, 7].

This work designs a WDM-NOMA system to enhance the data rate of the NOMA multiple access approach. NOMA and WDM-NOMA are evaluated using one colour, and four colours respectively. This is applied where one of the users is stationary and the other is mobile to study the impact of the users' positions on the achieved data rate. The paper is organized as follows: Section 2 explains the system modelling and simulation. The simulation results and discussion are presented in Section 3, then the conclusions are given in Section 4.

## 2. SYSTEM MODELLING AND SIMULATION

In the system model, an empty room with floor dimensions of 4 m × 8 m (length (Y) × width (X)), and ceiling height (Z) of 3 m was considered with a communication floor that is 1m high. Plaster walls usually reflect light in a form similar to a Lambertian function. Hence all walls and ceilings were modelled as Lambertian reflectors with reflectivity coefficient of 0.8. The room had no windows and the door was assumed to have a reflection coefficient similar to the that of the walls [8, 9]. The size of the surface elements decides the spatial resolution of the computation. The transmitter used is a Laser Diode (LD) with four colours (RYGB) providing four different channels of transmission.

The two scenarios evaluated NOMA and WDM-NOMA were considered in the same room configuration containing one access point and two users with a total power of 1W for the NOMA system and 0.8W, 0.5W, 0.3W and 0.3W for the red, yellow, green and blue respectively (to achieve the desired white colour) in NOMA-WDM. The access point is located on the ceiling at (2,5,3), the first user is located at a stationary point (1,2,1) while the second user is mobile along the Y axis, and moves from Y=2 to Y=8; i.e. from (2,2,1) to (2,8,1). Each scenario uses fair power allocation according to the channel strength of each user. The weaker the user's channel the more power it is allocated.

The line of sight (LOS) channel gain of the $k^{th}$ user is given by:

$$h_k = \begin{cases} \frac{(m+1)A_k}{2\pi d_k^2} \cos^m(\emptyset_k) T(\psi_k) g(\psi_k) \cos(\psi_k), & 0 \leq \psi_k \leq \Psi_k \\ 0, & \psi_k > \Psi_k \end{cases} \quad (1)$$

where $A_k$ is the area of the $k^{th}$ receiver detector; $d_k$ is the distance from the LD to the $k^{th}$ receiver; $\psi_k$ and $\emptyset_k$ are the incidence and the irradiance angles, respectively; $m = -1/\log_2\left(\cos\left(\frac{\Phi_1}{2}\right)\right)$ is the order of Lambertian emission; $\Psi_k$ and $\Phi_{1/2}$ are the $k^{th}$ receiver field of view (FOV) and the LD semi-angle at half power, respectively; $(\psi_k)$ is the optical filter gain; and $g(\psi_k) = \frac{n}{\sin(\Psi_k)^2}$ is the optical concentrator gain and $n$ is the refractive index.

The Signal to interference ratio (SINR) for the NOMA system is given by:

$$SINR_k = \frac{(a_k P_t R h_k \eta)^2}{(\sum_{\substack{i=1 \\ i \neq k}}^{K} a_i P_t R h_k \eta)^2 + (BN_0 + 2q(I_d + R P_{bn})B)} \quad (2)$$

where $a_k$ is the power allocation coefficient for the $k^{th}$ user and $P_t$ is the total electrical power, $R$ is the responsivity of the photodetector (Amps/Watt), $h_k$ is the optical channel gain, $\eta$ is the efficiency of the LD (Watts/Amp). For noise calculations, both thermal and shot noise were considered, where $N_0$ is the noise power density, $B$ is the receiver bandwidth and $q$, $I_d$ and $P_{bn}$ are the electron charge, the dark background current and received background optical power, respectively. The interference component is the sum of the power received from all other users' powers, where the number of users is denoted by K.

For the WDM-NOMA system, the SINR can be expanded and given by [19]:

$$SINR_c = \frac{R_c^2 (P_{c1} - P_{c0})^2}{\sigma_c^2 + I_c}, \quad c \in \{R, Y, G, B\} \quad (3)$$

where $R_c$ is the photodetector's responsivity associated with colour $c$. The received optical power associated with logic 1 is $P_{c1}$, while $P_{c0}$ is the received optical power associated with logic 0. The standard deviation of the total noise associated with the received data is $\sigma_c$ and has the same noise components used in the NOMA system mentioned in equation 2. $I_c$ is the interference for each colour received from multiple access points. Since only a single access point exists, the interference component from other colours is non-existent, $i.e. c=0 \ \forall \ C \in \{R, Y, G, B\}$.

The power allocation coefficient for each user is given by:

$$a_k = \frac{h_{(K-k+1)}}{\sum_{i=k+1}^{k} h_k} \quad (4)$$

and for the equal power allocation scenario, the power allocation coefficient is given by:

$$a_k = \frac{1}{K} \quad (5)$$

$$\sum_{i=k}^{K} a_k = 1$$

where the sum of the Power allocation fractions should always be '1' among all users.

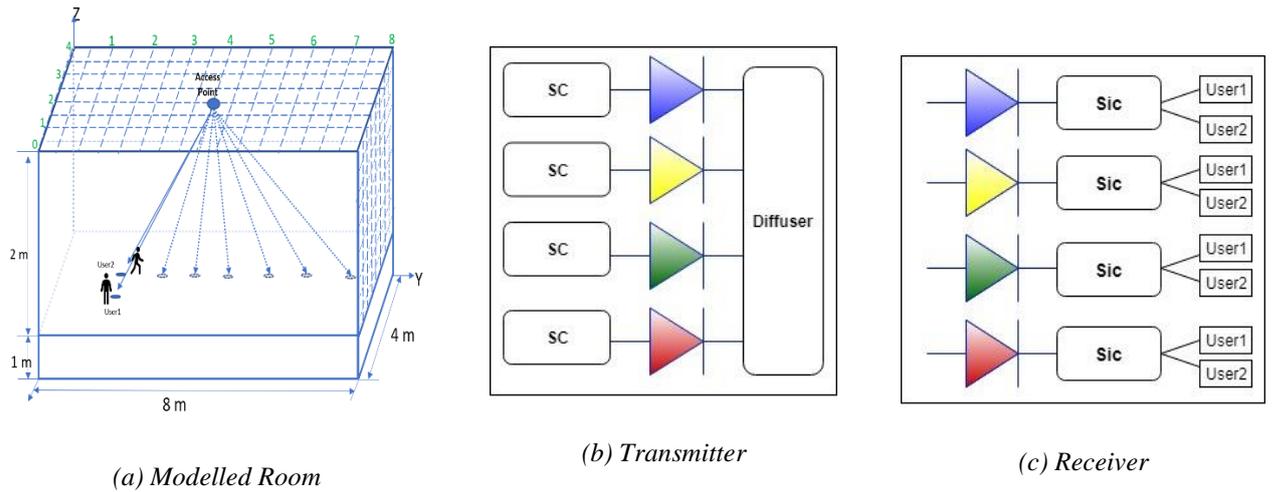

*(a) Modelled Room*   *(b) Transmitter*   *(c) Receiver*

*Figure 1. (a)Modelled Room (b) Transmitter (c) Receiver*

Each access point in the NOMA-WDM system transmits and receives signals using 4 different colours with a wide field of view. Each of these colours serves two users, as depicted in Figure 1. There are two main processes in NOMA systems, superposition code (SC) at the transmitter and successive interference cancellation (SIC) at the receiver's end to detect and cancel the signal sequentially from users with lower channels. Both processes are applied in each colour to provide the combined WDM and NOMA capabilities.

## 3. SIMULATION RESULTS AND DISCUSSION

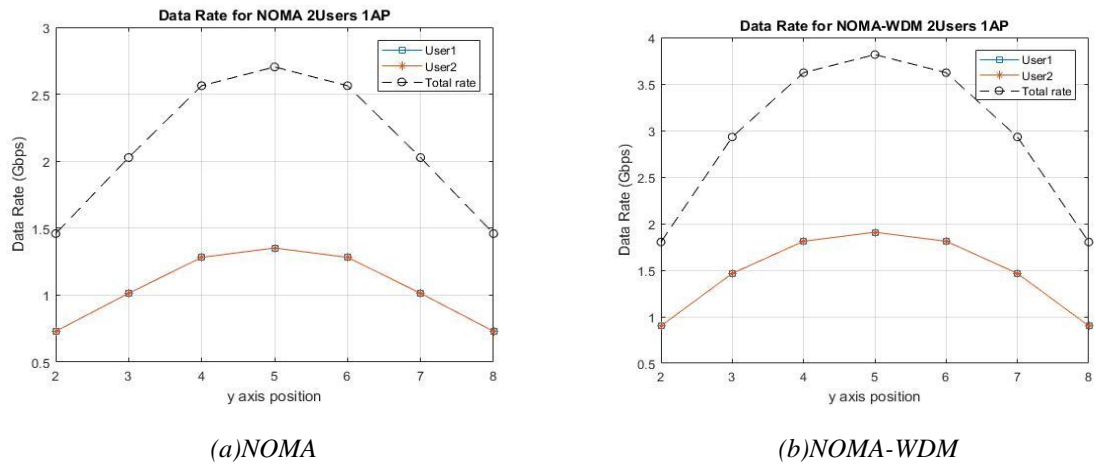

*(a)NOMA*   *(b)NOMA-WDM*

*Figure2.* Data rate vs the second user position for *(a) NOMA, (b) NOMA-WDM*

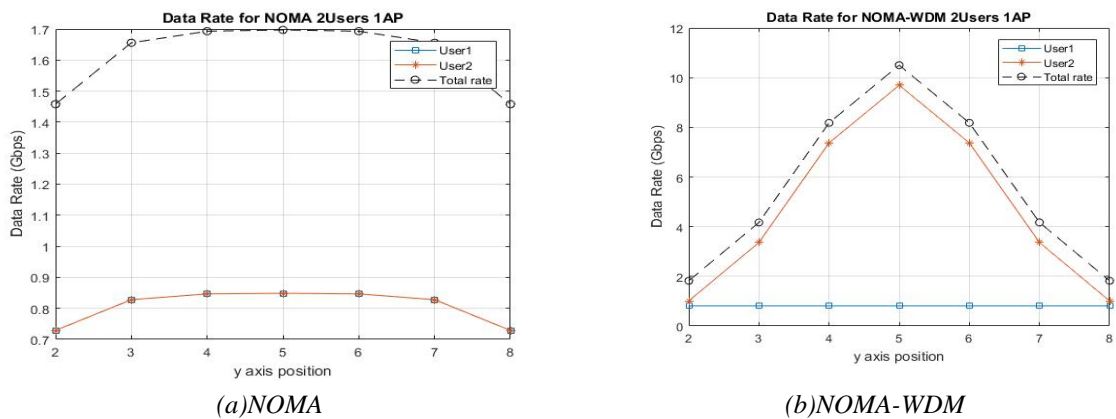

*(a)NOMA*   *(b)NOMA-WDM*

*Figure 3.* Data rate vs the second user position using equal power allocation for *(a) NOMA, (b) NOMA-WDM*

We compare NOMA-WDM with conventional NOMA where the parameters used are: $P_t = 1$ watt, $\eta = 1$, $\phi_{1/2} = 60°$, $A = 1 cm^2$; $\Psi = 60°$ for the NOMA system and $R_c = 0.4, 0.35, 0.3, 0.2$ for the red, yellow, green and blue colours respectively in the NOMA-WDM case, $T = 1$ ; n = 1.5 , $N_0 = 10^{-15} A^2/Hz$ and $L = 2m$. The horizontal distances between the users is about 0 - 3m. The optical power for the 4 Laser diodes are 0.8W, 0.5W, 0.3W and 0.3 for the red, yellow, green and blue colours respectively.

The results show that using WDM and NOMA results in higher gain in the rate for each user and the total sum rate for all users. The movement of the second user affects the power received by the first user and therefore its data rate, as depicted in Figure 2, and this consequently affects the power allocated to the first user. Therefore, the data rate of the first user that is stationary is affected by the mobility of the second user. The lowest data rate per user is 0.7 and the highest is 1.4 Gbps in the NOMA system and in the NOMA-WDM the minimum and the maximum data rates increased to 0.9 Gbps and 1.9Gbps, respectively. The location of the access point and the movement of the user affect the data rate, as the highest data rate achieved was when the mobile user was next to the access point was at the centre of the room.

Using the equal power allocation scheme results in much higher data rates but with more variations as shown in Figure 3. Data rates at the centre of the room can reach more than 10 Gbps. However, the sum rate is highly variable between users, as the mobile users get assigned the higher data rates. The total sum rate for the WDM-NOMA with equal power allocation has higher gain compared to the WDM-NOMA with conventional power allocation.

## 4. CONCLUSIONS

In this paper WDM-NOMA was developed to improve the total rate of the conventional NOMA system by using laser diodes with 4 colours on top of the NOMA multiple access scheme. The performance of NOMA-VLC systems and WDM-NOMA systems was compared considering different scenarios where the position of the users and the power allocation scheme were taken into consideration. The room simulated had one access point serving both users using four different colours. The simulation results proved that employing WDM-NOMA systems is better than NOMA systems, because it can provide enhanced service to the users by increasing the data rate per user.

### ACKNOWLEDGEMENTS
The authors would like to acknowledge funding from the Engineering and Physical Sciences Research Council (EPSRC), INTERNET (EP/H040536/1), STAR (EP/K016873/1) and TOWS (EP/S016570/1) projects. All data are provided in full in the results section of this paper.